\documentclass[dvips,twocolumn,aps,showpacs,amssymb,10pt]{revtex4}
\usepackage{graphicx}
\usepackage{dcolumn}
\usepackage{bm}
\usepackage{epsfig}
\usepackage{color}
\usepackage{longtable}
\usepackage{soul}
\usepackage{ulem}
\usepackage{amssymb,amsmath}
\usepackage{verbatim}
\everymath=\expandafter{\the\everymath\displaystyle}
\newcommand{\ds}{\displaystyle}

\newcommand{\Red}[1]{{\color{black}{#1}}}
\newcommand{\Blue}[1]{{\color{black}{#1}}}
\newcommand{\bra}[1]{\mathinner{\langle{#1}|}}
\newcommand{\ket}[1]{\mathinner{|{#1}\rangle}}

\newcommand{\redbra}[1]{\langle{#1}|\!|}
\newcommand{\redket}[1]{|\!|{#1}\rangle}

\newcommand{\dd}{\mathrm{d}}

\newcommand{\uvector}[1]{\hat{\bm{#1}}}
\renewcommand{\vec}[1]{\bm{#1}}

\newcommand{\sumint}[1]{\sum_{#1} \hspace{-0.5cm}\int\,}
\def\etal{{\it et al.}}                                        %
\begin{document}
\preprint{}
\title{Spin effects probed by Rayleigh X-ray scattering off hydrogenic ions} 
%
%

\author{L. Safari$^{1,}$\footnote{laleh.safari@oulu.fi}, P. Amaro$^{2,3}$, J. P. Santos $^{2}$, and F. Fratini $^{1,4,5}$} 

\affiliation{\it
$^1$ Department of Physics, University of Oulu, Box 3000, FI-90014 Oulu, Finland\\
$^2$ Centro de F\'{i}sica At\'{o}mica, Departamento de F\'{i}sica, Faculdade de Ci\^{e}ncias e Tecnologia, FCT, Universidade Nova de Lisboa, P-2829-516 Caparica, Portugal\\
$^3$ Physikalisches Institut, Universit\"{a}t Heidelberg, D-69120 Heidelberg, Germany\\
$^4$ Departamento de F\'isica, Universidade Federal de Minas Gerais, 30123-970 Belo Horizonte, Brazil\\
$^5$ Institut N\'eel-CNRS, BP 166, 25 rue des Martyrs, 38042 Grenoble Cedex 9, France} 
\date{\today \\[0.3cm]}%
%
%
%
\begin{abstract}
We study the polarization characteristics of x-ray photons scattered by hydrogenic atoms, based on the Dirac equation and second-order perturbation theory. The relativistic states used in calculations are obtained using the finite basis set method and expressed in terms of $B$-splines and $B$-polynomials. We derive general analytical expressions for the polarization-dependent total cross sections, which are applicable to any atom and ion, and evaluate them separately for linear and circular polarization of photons. In particular, detailed calculations are performed for the integrated Stokes parameters of the scattered light for hydrogen as well as hydrogenlike neon and argon. Analyzing such integrated Stokes parameters, special attention is given to the electron-photon spin-spin interaction, which mostly stems from the magnetic-dipole contribution of the electron-photon interaction. Subsequently, we find an energy window for the selected targets in which such spin-spin interactions can be probed.
\end{abstract}
\maketitle
%
%
%
%
%
\section{Introduction}\label{sec:intoduc}

Photonic spin, which is defined by its polarization, is one of the fundamental characteristics of light. Understanding the effects that spin induces to physical phenomena is of great importance in many branches of applied science and technology like quantum communications, astronomy, chemistry, electronics and so on. \Blue{Owing to the recent advances in x-ray polarization sensitive detectors \cite{G.Weber:2010A,U.Spillmsnn:2008,D.Protic:2005}, tunable polarization and orbital angular momentum (OAM) free-electron lasers (FELs) \cite{C.Spezzani:11,Hemsing:2013,Hemsing:2014,F.Fratini:2014} as well as synchrotron radiation sources \cite{F.Smend:87}, an increasing demand for accurate theoretical prediction on polarization-dependent atomic phenomena is arising. On the other hand, due to the recent progresses in storage rings and trapping techniques the scattering of light by ionic targets is of special interest nowadays since it allows us to obtain information about electron-photon interactions as well as atomic structures in the presence of strong Coulomb field. This also calls \Red{for} accurate theoretical predictions in order to complement the experimental data.

The mentioned theoretical predictions address fundamental level issues, which are, consequently, not only essential for the new generations of experiments but also are of high importance for other fields. For instance, very recently, a measurement of the angle-differential cross section for Rayleigh scattering off some selected neutral atoms with fully linearly polarized incident x-rays has been performed in the PETRA III synchrotron at DESY \cite{Alex:2013}. Soon after, there has been a theoretical work focused on analyzing angle-differential cross sections for Rayleigh scattering of linearly polarized x-rays by some hydrogenlike ions (see. \cite{A.Surzhykov:2013} and references therein). On the other hand, polarization-dependent {\it total cross section} (PDTCS) for scattered light can be measured via 4$\pi$ detectors. Furthermore, highly energetic photons mostly scatter forwardly. Therefore, PDTCSs may be measured up to a very good approximation by current techniques of measuring angle-differential cross sections.\\
Polarization analysis in scattering of highly energetic photons is also of great interest in astrophysics and cosmology. Related to PDTCS,} the polarization power spectra in Rayleigh scattering has been regarded as a promising tool for investigating Cosmic Microwave Background (CMB) anisotropies from recombination processes \cite{A.Lewis:2013}. \Blue{Moreover, the PRISM (Polarized Radiation Imaging and Spectroscopy Mission) collaboration aims to measure temperature signal and polarization from Rayleigh scattering with high accuracy in order to set \Red{constraints} in the early recombination history \cite{PRISM:2014}. Polarized light in the interstellar medium can be originated by the scattering of light by inhomogeneously populated hydrogen and helium atoms. The processes that generate inhomogeneous atomic population are often the presence of a stellar magnetic field (Zeeman and Paschen-Back effects) or anisotropic radiation pumping. Specifically, a magnetic field that is inclined with respect to the symmetry axis of the pumping radiation can produce significant linear polarization signals in spectral lines (the so-called Hanle effect) \cite{J.Trujillo:06,E.Landi:04}. Polarization analysis of starlight is therefore essential for providing reliable information on stellar magnetic fields \cite{J.O.Stenflo:06,A.J.Dean:08}.}

When light travels through the matter it partially loses its polarization due to atom-photon interactions. In this article we address the investigation of such \Blue{interactions. Specifically, in Sec. \ref{sec:Theory}, we derive analytical expressions for PDTCS, which are valid for scattering by hydrogenic as well as by many-electron systems. The main advantage of such analytical expressions is that allows us to perform faster and more accurate calculations. We then evaluate those expressions for two experimental scenarios that uses light  circularly (Sec. \ref{subsec:Pol-scenario}) and linearly (Sec. \ref{subsubsec:Lin-scenario}) polarized. By making use of Wigner-Racah algebra, we write those expressions in terms of angular parts and reduced matrix elements. In Sec. \ref{sec:Comput}, the fully relativistic numerical evaluation of the reduced matrix elements is carried out through the use of finite basis sets for the scattering by hydrogen as well as hydrogenlike neon ($\textrm{Ne}^{9+}$) and argon ($\textrm{Ar}^{17+}$). In Sec. \ref{sec:ResultsDiscussion}, particular attention is paid to the analysis of integrated Stokes parameters (ISPs) of the scattered light as a powerful tool for probing effects of spin-spin interactions during the scattering events. Such effects are found to be within $10^{-3}$ \% to 2 \% from hydrogen to hydrogenlike argon. The investigated energy range for all targets lies above their $1s$ ionization thresholds. It has been recently showed by us that, within that energy range, the finite basis set approach to Rayleigh scattering gives good agreement with NIST (National Institute of Standards and Technology) data values and other calculations \cite{L.Safari:12_B,L.Safari:12}. Finally, a summary of our findings and links to future prospects are given in Sec. \ref{sec:SumConcl}.}

%
%
%
%
%
\section{Theory}
\label{sec:Theory}
\subsection{Geometry of scattering event}\label{sec:Geometry}

\begin{figure}[t]
\includegraphics[scale=0.6]{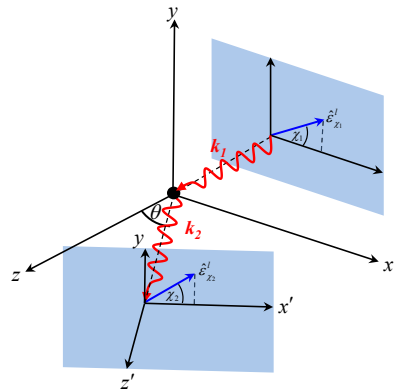}
\caption{(Color online) The adopted geometry for the scattering event. The polar angle $\theta$ uniquely defines the direction of the scattered photon in the scattering plane. The angle $\chi_1$ ($\chi_2$) parametrizes the linear polarization of the incident (scattered) photon. The hydrogenic atom is located at the origin of the coordinate axes $xyz$.}
\label{Rayleigh geometry}
\end{figure}

%
%
%
%
The geometry under which our studies are conducted is displayed in Fig.~\ref{Rayleigh geometry}. Hydrogenic atom irradiated by energetic light is considered in the ground state. Incident (scattered) photons have energies $E_{\gamma_{1 (2)}}= \hbar \omega_{1 (2)}$, propagation \Red{vectors} $\vec{k}_{1 (2)}$ and polarization unit vectors $\uvector{\bm{\epsilon}}_{1 (2)}$, where $\hbar$ is the reduced Planck constant. 
For both, the incident and scattered photons, $\uvector\epsilon$ is assumed to be either linear ($\uvector\epsilon_\chi^l$) or circular ($\uvector\epsilon_\lambda^c$). The polarization angle $0 \le \chi \le \pi$ and the helicity $\lambda = \pm 1$ are the variables used to parametrize the linear and circular polarization states, respectively. In such a notation, $\lambda = +1$ describes right-handed and $\lambda = -1$ left-handed polarized photons, respectively. The (definition of the) polarization angles are also displayed in Fig.~\ref{Rayleigh geometry} if incident and scattered photons polarizations are taken to be linear. The direction  of the incident photon ($\vec{k}_1$) defines the quatization ($z$) axis. This selection of the quantization axis simplifies the multipole expansion of the electron-photon interaction operator. The scattered photon propagates along the direction $\vec{k}_2$ at angle $\theta$ with respect to the $z$ axis. The scattering plane ($xz$) is defined by the incident and scattered photon directions. 
%
%

%
%
%
%
%
\subsection{Second-order scattering transition amplitude}
\label{subsec:Transition ampl}

The second-order transition amplitude for Rayleigh (and Raman) scattering is given by \cite{L.Safari:12,L.Safari:12_B,Filippo:11,Akhiezer:65}
\begin{eqnarray}
\label{Mamplitude}
            \mathcal{M}_{if}(\uvector\epsilon_1,\uvector\epsilon_2)& =&
              \sumint{\nu} \frac{\bra{f}\mathcal{R}^{\dagger}(\vec k_2,\uvector\epsilon_2)
                \ket{\nu}\bra{\nu}\mathcal{R}(\vec k_1,\uvector\epsilon_1)\ket{i}}{\omega_{\nu i}-\omega_1}  \nonumber\\
           & & \hspace{-1cm} +  \sumint{\nu}\frac{\bra{f}\mathcal{R}(\vec k_1,\uvector\epsilon_1)
                \ket{\nu}\bra{\nu}\mathcal{R}^{\dagger}(\vec k_2,\uvector\epsilon_2)\ket{i}}{\omega_{\nu i}+\omega_2} \, ,
\end{eqnarray}
where $\omega_{\nu i} = (E_{\nu}-E_i)/\hbar$ is the transition frequency between states $\ket{\nu}$ and $\ket i$. Here, the transition operator $\mathcal{R}(\vec k_{1(2)},\uvector\epsilon_{1(2)})$ describes the relativistic interaction between the bound electron and the incident (scattered) photon. 
In the Coulomb gauge, the explicit expression of this transition operator is
\begin{eqnarray}
\label{R operator}
    \mathcal{R}(\vec k,\uvector\epsilon)=\bm{\alpha}\cdot\uvector{\epsilon} 
    \, e^{i\vec k\cdot \vec r} ~,
\end{eqnarray}
where $\bm{\alpha}$ is the vector of Dirac matrices. The initial $\ket i$ and final $\ket f$ atomic states have well-defined angular momentum $j$, angular momentum projection $m_j$ and parity $(-1)^l$, where $l$ is the orbital angular momentum of the larger component of the Dirac spinor.
In the following, such states will be labeled by $\ket{i}\equiv\ket{\beta_i,j_i,m_{j_i}}$ and 
$\ket{f}\equiv\ket{\beta_f,j_f,m_{j_f}}$, where $\beta$ is a collective label used to denote all the additional quantum numbers needed to specify the atomic states but for $j$ and $m_j$. For hydrogenic ions, specifically, $\beta$ refers to the principal quantum number $n$ and the orbital angular momentum quantum number $l$. Since Rayleigh scattering is an elastic process, the initial and final states coincide, $\ket{i}\,=\,\ket{f}$ and consequently the energy of incident and scattered photons becomes equal ($ E_{\gamma_1}=E_{\gamma_2}\equiv E_{\gamma}$).

In this work, we shall separately consider the two experimental scenarios corresponding to circularly and linearly polarized light. In the following subsections, we will show how PDTCs and ISPs are defined in terms of the transition amplitude \eqref{Mamplitude}, and evaluate them separately for each scenario. Moreover, in both scenarios, is assumed that the initial and final atomic states remain unobserved as typical in most experiments.

%
%
%
%
\subsection{Circular polarization scenario}
\label{subsec:Pol-scenario} 

We here consider an experimental scenario in which the incident light is circularly polarized and the polarization of the scattered light is measured in the circular base (i.e., the helicity of the scattered photon is measured). In this scenario, the angular distribution function (i.e., the angle-differential cross section) can be
written in SI units as \cite{L.Safari:12}
\begin{equation}
\begin{array}{lcl}
\label{dsig/domega_cc}
\frac{\dd \sigma^{cc}}{\dd \Omega}(\lambda_1,\lambda_2)
         &=&\frac{\alpha^2c^2}{(2j_i+1)}\sum_{\substack{m_{j_i}m_{j_f}}}
         \Big|\mathcal{M}^{cc}(\lambda_1,\lambda_2)\Big|^2 \,,
\end{array}
\end{equation}
where $\alpha$ is the electromagnetic coupling constant, $c$ the speed of light in vacuum, and we considered $\mathcal{M}^{cc}(\lambda_1,\lambda_2)\equiv\mathcal{M}_{if}(\uvector\epsilon_{\lambda_1}^c,\uvector\epsilon_{\lambda_2}^c)$.\\
To further facilitate the analysis of the angular and polarization properties of the transition amplitude $\mathcal{M}^{cc}(\lambda_1,\lambda_2)$, we expand $\uvector\epsilon_\lambda^c e^{i\vec k\cdot \vec r}$ in terms of spherical tensors with well-defined angular momentum properties \cite{M.E.Rose:57}
\begin{eqnarray}
\label{multi-pole decomposition final}
              {\uvector\epsilon}_{\lambda}^c e^{i\vec k\cdot \vec r} &=&
              \sqrt{2\pi}\sum^{+\infty}_{L=1}\sum^{L}_{M=-L}\sum_{p=0,1}  
                  i^L[L]^{1/2} (i\lambda)^p\,\bm{a}^p_{LM}(k,\vec r) \nonumber\\
              & & \times\, D^L_{M\lambda}(\varphi_k,\theta_k,0)    ~,
\end{eqnarray}
where $[L_1,L_2,...,L_n]=(2L_1+1)(2L_2+1)...(2L_n+1)$, while the spherical tensor $\bm{a}^p_{LM}(k,\vec r)$ refers to the magnetic ($p=0$) and electric ($p=1$) multipoles. On the other hand, $k$ refers to $|\vec k|$. Each term $\bm{a}^p_{LM}(k,\vec r)$ has angular momentum $L$, angular momentum projection $M$ and parity $(-1)^{L+1+p}$. Expressions for $\bm{a}^p_{LM}(k,\vec r)$ can be found in \cite{L.Safari:12}. Combining Eqs.~\eqref{Mamplitude} and \eqref{multi-pole decomposition final}, and making use of the Wigner-Eckart theorem \cite{Sakurai:94}, the transition amplitude can be written as
\begin{widetext}
\begin{equation}
\label{eq:M}
\begin{array}{lcl}
       \ds\mathcal{M}^{cc}(\lambda_1,\lambda_2)&=&
        2\pi\sum_{\substack{L_1L_2\\M_2}}\sum_{p_1p_2}(+i)^{L_1-L_2+p_1+p_2}
        [L_1,L_2]^{1/2}(\lambda_1)^{p_1}(\lambda_2)^{p_2}
        d^{L_2}_{M_2-\lambda_2}(\theta)
\\[0.8cm]
   &&   \times\;\sum_{j_\nu}
        (-1)^{-j_{\nu}}\frac{1}{(2j_{\nu}+1)^{1/2}}
        \Big(\Theta^{j_{\nu}}(1,2)S^{j_{\nu}}(1,2)
        +\Theta^{j_{\nu}}(2,1)S^{j_{\nu}}(2,1)
        \Big)~,
\end{array}
\end{equation}
where the reduced (second-order) matrix element is given by
\begin{equation}
\label{eq:S_J}
 S^{j_\nu}(1,2)= 
\sum_{\beta_{\nu}}\frac{\redbra{\beta_i,j_i}\bm\alpha\cdot\bm a_{L_1}^{p_1}(k_1,\vec r) \redket{\beta_{\nu},j_{\nu}}
\redbra{\beta_{\nu},j_{\nu}}\bm\alpha\cdot\bm a_{L_2}^{p_2}(k_2,\vec r)\redket{\beta_{i},j_{i}}}
{\omega_{\nu i}+\omega_2}~,
\end{equation}
\end{widetext}
and $S^{j_\nu}(2,1)$ is obtained from \eqref{eq:S_J} by i) interchanging the label 1 with 2 and ii) replacing the positive sign in the denominator with a negative sign. The reduced amplitude $\redbra{\beta_\nu,j_\nu}\bm\alpha\cdot\bm a_{L}^{p}(k,\vec r) \redket{\beta_{i},j_{i}}$ does not depend on the angle $\theta$ and is zero unless $\Pi_i\,(-1)^{L+p+1}=\Pi_\nu$, 
where $\Pi_{i(\nu)}$ is the parity of initial (intermediate) hydrogenic state and has the well-known form of $(-1)^{l_{i(\nu)}}$. Following the notation used in \cite{S.P.Goldman:81, L.Safari:12}, in Eq. \eqref{eq:M} we have furthermore defined  
\begin{eqnarray}
\label{eq:ThetaS}
\Theta^{j_{\nu}}(1,2)&=&\sum_{m_{j_\nu}}(-1)^{m_{j_f}+m_{j_{\nu}}}(2j_{\nu}+1)^{1/2} \\\nonumber
     &\times& \left(\begin{array}{ccc}
     j_f&L_1&j_{\nu}\\
     -m_{j_f}&\lambda_1&m_{j_{\nu}}
    \end{array}
\right)\left(
\begin{array}{ccc}
j_{\nu}&L_2&j_{i}\\
-m_{j_{\nu}}&M_2&m_{j_{i}}
\end{array}
\right)~,
\end{eqnarray}
where $\Theta^{j_{\nu}}(2,1)$ is obtained from Eq.~\eqref{eq:ThetaS} by replacing $L_1\leftrightarrow L_2$ and $\lambda_1\leftrightarrow M_2$. From Eqs. \eqref{dsig/domega_cc}-\eqref{eq:S_J}, and by integrating the angle-differential cross section \eqref{dsig/domega_cc} over the angle $\theta$, the PDTCS can be written in SI units as
\begin{equation}
\label{sigma_cc}
\sigma^{cc}(\lambda_1,\lambda_2)=
       \frac{\alpha^2c^2}{(2j_i+1)}\sum_{\substack{m_{j_i}m_{j_f}}}
       \sum_{L_2M_2}\Big|
       \tilde{\mathcal{M}}^{cc}(\lambda_1,\lambda_2)
       \Big|^2~,
\end{equation}
where 
\begin{widetext}
\begin{equation}
\label{Mtild}
\begin{array}{lcl}
\tilde{\mathcal{M}}^{cc}(\lambda_1,\lambda_2)
&=& 4\pi\sqrt{\pi}\sum_{L_1}\sum_{p_1p_2}(+i)^{L_1+p_1+p_2}[L_1]^{1/2}
    (\lambda_1)^{p_1}(\lambda_2)^{p_2}
\\[0.6cm]
&&   \times\;\sum_{j_\nu}
    (-1)^{-j_{\nu}}\frac{1}{(2j_{\nu}+1)^{1/2}}
    \Big(\Theta^{j_{\nu}}(1,2)S^{j_{\nu}}(1,2)
    +\Theta^{j_{\nu}}(2,1)S^{j_{\nu}}(2,1)\Big)~.
\end{array}
\end{equation}
In order to obtain Eq. \eqref{sigma_cc}, the integral result
\begin{equation}
\begin{array}{c}
\int_0^{2\pi}\dd\alpha\int_0^\pi \dd\beta\sin\beta \;
D_{m_1\,b}^{J_1\,*}(\alpha,\beta,0)\,D_{m_2\,b}^{J_2}(\alpha,\beta,0)=
\frac{4\pi}{2J_1+1}\,\delta_{m_1, m_2}\delta_{J_1, J_2} 
\end{array}
\label{eq:integralSimpl}
\end{equation}
\end{widetext}
has been used. Although, from Eq. \eqref{sigma_cc}, the quantity $\tilde{\mathcal{M}}^{cc}$ may seem a quantum-mechanical amplitude, we underline that, rigorously, it is {\it not}. The quantity $\tilde{\mathcal{M}}^{cc}$ has been obtained after having \Red{integrated} over the scattered photon direction, and, therefore, does not fulfill the superposition rule that applies to any quantum-mechanical amplitude. 

The only Stokes parameter that is relevant for circularly polarized light is the third one and is defined as
\begin{equation}
\label{eq:intP3_cc}
\tilde{P}_C^{cc}\equiv\tilde{P}_3^{cc}(\lambda_1, E_\gamma)=
           \frac{\sigma^{cc}(\lambda_1,+1)
          -\sigma^{cc}(\lambda_1,-1)}
          {\sigma^{cc}(\lambda_1,+1)
          +\sigma^{cc}(\lambda_1,-1)} ~.
\end{equation}
The quantity $\tilde{P}_3^{cc}(\lambda_1, E_\gamma)$ does not depend on the scattering angle $\theta$ but, rather, depends only on the energy ($E_\gamma$) and on the circular polarization ($\lambda_1$) of the incident photon.
It measures the circular polarization of the scattered light for events in which the incident light is circularly polarized. It ranges from -1 to +1 and the value -1 (+1) corresponds to events in which scattered light is fully left(right)-handed polarized.

For the unpolarized incident light case, we can obtain the total cross sections by averaging Eq. (\ref{sigma_cc}) over the polarizations of the incident light \cite{F.Fratini-Hyrapetyan:11}. 
Upon doing this, we obtain
\begin{equation}
\label{sigma_unpc}
\sigma^{uc}(\lambda_2)=
       \frac{\alpha^2c^2}{2(2j_i+1)}\sum_{\substack{m_{j_i}m_{j_f}\\\lambda_1}}
       \sum_{L_2M_2}\Big|\tilde{\mathcal{M}}^{cc}(\lambda_1,\lambda_2)\Big|^2~.
\end{equation}
Correspondingly, the third ISP for this case takes the form
\begin{equation}
\label{eq:intP3_uc}
\tilde{P}_3^{uc}(E_\gamma)=
           \frac{\sigma^{uc}(+1)
          -\sigma^{uc}(-1)}
          {\sigma^{uc}(+1)
          +\sigma^{uc}(-1)} ~,
\end{equation}
where $\tilde{P}_3^{uc}$ measures the circular polarization of the scattered light for events in which the incident light is unpolarized.

%
%
%
%
%
\subsection{Linear polarization scenario}
\label{subsubsec:Lin-scenario} 

We will address in this section an experimental scenario in which the incident light is linearly polarized, and the polarization of the scattered light is measured in the linear base. The angular distribution is given by \cite{L.Safari:12_B}
\begin{equation}
\label{dsig/domega_ll}\frac{\dd \sigma^{ll}}{\dd \Omega}(\chi_1,\chi_2)=\frac{\alpha^2c^2}{(2j_i+1)}\sum_{\substack{m_{j_i}m_{j_f}}}\Big|\mathcal{M}^{ll}(\chi_1,\chi_2)\Big|^2 ~,
\end{equation}
where we considered $\mathcal{M}^{ll}(\chi_1,\chi_2)\equiv\mathcal{M}_{if}(\uvector\epsilon^l_{\chi_1},\uvector\epsilon^l_{\chi_2})$. The amplitude for linear polarization scenario is linked to the amplitude for circular polarization scenario as \cite{L.Safari:12_B}
\begin{eqnarray}
\label{eq:ll ampl}
      \mathcal{M}^{ll}(\chi_1,\chi_2)
      =\frac{1}{2}\sum_{\lambda_1\lambda_2}e^{-i\lambda_1 \chi_1}e^{i\lambda_2 \chi_2}
      \mathcal{M}^{cc}(\lambda_1,\lambda_2)~.
\end{eqnarray}
Upon integrating $\dd\sigma^{ll}/\dd\Omega$ over the scattered photon direction, we get the total cross section as
\begin{equation}
\label{sigma_ll}
\sigma^{ll}(\chi_1, \chi_2)=\int d\Omega \frac{d \sigma^{ll}}{d \Omega}=
       \frac{\alpha^2c^2}{(2j_i+1)}\sum_{\substack{m_{j_i}m_{j_f}}}
        \mathcal{L}^{ll}(\chi_1,\chi_2) ~,
\end{equation}
where 
\begin{widetext}
\begin{eqnarray}
\mathcal{L}^{ll}(\chi_1,\chi_2)&=& (2\pi)
\sum_{\substack{L_1 L_2 \\ L_1' L_2'}}
\sum_{\substack{p_1 p_2 \\ p_1' p_2'}}
\sum_{\substack{\lambda_1 \lambda_2 \\ \lambda_1' \lambda_2'}}
\sum_{\substack{m m'\\M_2}}
\left(
V_{\lambda_1 \lambda_2 M_2}^{L_1 L_2 p_1 p_2}\,e^{i\frac{\pi}{2}\lambda_2}\,h^{L_2}_{M_2 m \lambda_2}
\right) 
\left(
V_{\lambda_1' \lambda_2' M_2}^{L_1' L_2' p_1' p_2'}\,e^{i\frac{\pi}{2}\lambda_2'}\,h^{L_2'}_{M_2 m' \lambda_2'}
\right)^*
 \; f(m, m')\, , \\ \nonumber
V_{\lambda_1 \lambda_2 M_2}^{L_1 L_2 p_1 p_2}&=&
\pi e^{i(\lambda_2\chi_2-\lambda_1\chi_1)} \, (i)^{L_1-L_2+p_1+p_2}\, [L_1,L_2]^{1/2}\,
(\lambda_1)^{p_1}(\lambda_2)^{p_2} \\ \nonumber
 &\times&\;\sum_{j_\nu}
    (-1)^{-j_{\nu}}\frac{1}{(2j_{\nu}+1)^{1/2}}
    \Big(\Theta^{j_{\nu}}(1,2)S^{j_{\nu}}(1,2)
    +\Theta^{j_{\nu}}(2,1)S^{j_{\nu}}(2,1)\Big) ~,
\end{eqnarray}		
\begin{eqnarray}
f(n,m)&=& 2\delta_{n,m} + i\frac{\pi}{2}\Big( \delta_{n,m+1} - \delta_{n,m-1} \Big)\, ,~~\textrm{and}~~
h^{J}_{n m k}\,=\,
d^{J}_{n\,m}(-\pi/2)\,
d^{J}_{m\,k}(\pi/2)~.
\end{eqnarray}
In order to obtain Eq. \eqref{sigma_ll}, the integral
\begin{equation}
\begin{array}{c}
\int_0^{2\pi}\dd\alpha\int_0^\pi \dd\beta\sin\beta \;
D_{m_1\,b_1}^{J_1\,*}(\alpha,\beta,0)\,D_{m_2\,b_2}^{J_2}(\alpha,\beta,0)=
(2\pi)\,\delta_{m_1,m_2}\,e^{-i\frac{\pi}{2}(m_1-m_2)}\,e^{+i\frac{\pi}{2}(b_1-b_2)}\\
\times\sum_{m_3=-J_1}^{J_1}\sum_{m_5=-J_2}^{J_2} \,f(m_3,m_5) \,h^{J_1}_{m_1 m_3 b_1}\,h^{J_2}_{m_2 m_5 b_2}
\end{array}
\label{eq:integral}
\end{equation}
\end{widetext}
has been used, where $J_1$ and $J_2$ are integers.
Equation \eqref{eq:integral} can be in turn obtained by decomposing a rotation of a certain angle $\beta$ around the $y$ axis into a rotation of the same angle around the $z$ axis, followed and preceded by (non-commutative) rotations of angles $\pm 90^\circ$ around $y$ and $z$ axes \cite{Edmond:eq.4.5.1}. If $b_1=b_2\equiv b$, the integral \eqref{eq:integral} simplifies into Eq.~\eqref{eq:integralSimpl}. \Blue{As mentioned before, Eq. \eqref{sigma_ll} allows us to calculate the total cross section faster and more accurately in comparison with numerical integrations.}

The integrated degree of linear polarization is given by
\begin{equation}
\label{eq:DegreeCircLin}
\tilde{P}_L=\sqrt{(\tilde{P}_1^{ll})^2+(\tilde{P}_2^{ll})^2}~,
\end{equation}
where the first and the second ISPs are defined as
\begin{equation}
\label{eq:int_P1_ll}
\begin{array}{l}
\tilde{P}_1^{ll}(\chi_1, E_\gamma)=
           \frac{\sigma^{ll}(\chi_1,0^\circ)
          -\sigma^{ll}(\chi_1,90^\circ)}
          {\sigma^{ll}(\chi_1,0^\circ)
          +\sigma^{ll}(\chi_1,90^\circ)} ~,\\ [0.4cm]
\tilde{P}_2^{ll}(\chi_1, E_\gamma)=
           \frac{\sigma^{ll}(\chi_1,45^\circ)
          -\sigma^{ll}(\chi_1,135^\circ)}
          {\sigma^{ll}(\chi_1,45^\circ)
          +\sigma^{ll}(\chi_1,135^\circ)} ~,
\end{array}
\end{equation}
respectively. $\tilde{P}_{1(2)}^{ll}$ ranges from -1 to +1 and measures the linear polarization of the scattered light for events in which the incident light is linearly polarized. More specifically, $\tilde{P}_{1(2)}^{ll}$ measures the linear polarization along the chosen pair of axes defined by the angles $\chi_1=0^\circ$ ($45^\circ$), $\chi_2=90^\circ$ ($135^\circ$).

In the case the incident light is unpolarized, while the scattered light is measured in the linear base, we can obtain the total cross sections by averaging Eq.~(\ref{sigma_ll}) over the polarizations of the incident light as \cite{F.Fratini-Hyrapetyan:11}
\begin{equation}
\label{sigma_unpl}
\sigma^{ul}(\chi_2)=
       \frac{\alpha^2c^2}{2(2j_i+1)}\sum_{\substack{m_{j_i}m_{j_f}\\\chi_1}}
       \mathcal{L}^{ll}(\chi_1,\chi_2)~.
\end{equation}
The first and the second ISPs are thus defined as
\begin{equation}
\label{eq:intP1_ul}
\begin{array}{l}
\tilde{P}_1^{ul}(E_\gamma)=
           \frac{\sigma^{ul}(0^\circ)
          -\sigma^{ul}(90^\circ)}
          {\sigma^{ul}(0^\circ)
          +\sigma^{ul}(90^\circ)} ~,\\ [0.4cm]
\tilde{P}_2^{ul}(E_\gamma)=
           \frac{\sigma^{ul}(45^\circ)
          -\sigma^{ul}(135^\circ)}
          {\sigma^{ul}(45^\circ)
          +\sigma^{ul}(135^\circ)}~,					
\end{array}					
\end{equation}
where $\tilde{P}_{1(2)}^{ul}$ measures the linear polarization of the scattered light for events in which the incident light in unpolarized.

%
%
%
%
%
%
%
\section{Numerical calculations}
\label{sec:Comput}

The most difficult task of obtaining the transition amplitude \eqref{Mamplitude} is the calculation of S-function in Eq. \eqref{eq:S_J}. In the present work, we calculate the S-function by adopting the \textit{finite basis set} method, using B splines and B polynomials as finite basis sets \cite{saj1996,jbs1998,ZatBar2008,FroZat2009,sfi1998,P.M.D.G.Amaro}. The parameters of the B splines basis set are the radius of the cavity ($R_{\rm{bs}}$), the number of B splines ($n_{\rm{bs}}$) and their degree ($k$). As for the B polynomials, the parameters are the radius of the cavity ($R_{\rm{bp}}$) and the number of B polynomials ($n_{\rm{bp}}$) (the degree of the B polynomials is $n_{\rm{bp}}-1$). The parameters used in both basis sets were optimized in order to obtain stability and an agreement of six digits between the results produced by the two approaches. The optimal used parameters are:  $R_{\rm{bs}}=60$~a.u., $n_{\rm{bs}}=60$, $k=9$, $R_{\rm{bp}}=50$~a.u. and $n_{\rm{bp}}=40$. There has been also alternative numerical methods such as the Coulomb-Green function for calculating second-order transition amplitudes in the literature (see Refs. \cite{F.Fratini:11_2,F.Fratini-S.Trotsenko:11,P.Amaro:2012,A.Surzhykov:10,PolPol,L.Safari:2014}).

%
%
%
%
%
\section{Results and discussion}
\label{sec:ResultsDiscussion}

In this section, \Blue{we present the results for the integrated degrees of circular and linear polarization of the scattered light by hydrogen as well as hydrogenlike neon ($\textrm{Ne}^{9+}$) and argon ($\textrm{Ar}^{17+}$). Results have been obtained by using the relations presented in Secs. \eqref{subsec:Pol-scenario} and \eqref{subsubsec:Lin-scenario} and by implementing the computation technique presented in Sec. \ref{sec:Comput}.}

%
%
%
%
%
\subsection{Circular polarization scenario}
\label{Circ}
\begin{figure}[t]
\includegraphics[scale=0.73]{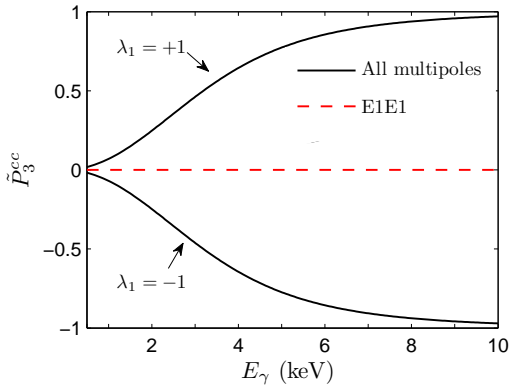}
\caption{(Color online) 
Third ISP of the scattered light by hydrogen when incident light is circularly polarized ($\tilde{P}_3^{cc}$), as a function of the photon energy.
The curves corresponding to positive ($\lambda_1=+1$) and negative ($\lambda_1=-1$) helicity of the incident photon are displayed.
Results are calculated with the account of all photon multipoles (solid-black line) and within the electric dipole approximation (dashed-red line).}
\label{fig:4}
\end{figure}
Here we consider the circular polarization scenario as described in Sec. \ref{subsec:Pol-scenario}. Fig.~\ref{fig:4} displays the third ISP, $\tilde{P}_3^{cc}$, for scattering by hydrogen, as a function of the photon energy $E_\gamma$ within the energy range 0.5 to 10 keV and for the two initial photon helicities $\lambda_1=\pm1$. Results are shown by taking into account only the contribution of electric-dipole (E1E1) multipole (dashed-red-line) and the contribution of all multipoles of the electron-photon interaction (solid-black-line). As seen in the figure, for the E1E1 curve, $\tilde{P}_3^{cc}$ is zero for all energy values. This can be understood in view of the symmetric shape of angular distribution for E1E1 term \cite{L.Safari:12,L.Safari:12_B}: The photon is scattered at small angles (in which case helicity is conserved) or large angles (in which case helicity is flipped) with the same probability. This naturally determines a vanishing third ISP. A similar discussion also applies to the higher multipoles: For low photon energies, the E1E1 multipole in the photonic interaction operator has the most contribution and, consequently, the third ISP turns out to be around zero in either case the photon initial helicity being positive or negative; For high photon energies, forward scattering is strongly preferred to backward scattering \cite{L.Safari:12}, which results in the conservation of the photon helicity during the scattering process. \Blue{Although scaled at higher photon energies, plots of $\tilde{P}_3^{cc}$ for scattering by targets $\textrm{Ne}^{9+}$ and $\textrm{Ar}^{17+}$, which we do not show here for the sake of simplicity, follow the same behavior as for hydrogen. Therefore, the foregoing discussion also applies to them.}

If incident light is unpolarized, $\tilde{P}_3^{uc}$ vanishes for all energies \Blue{and for all targets.} This result can be traced back to conservation of parity \cite{L.Safari:12}.

%
%
%
%
%
\subsection{Linear polarization scenario}
\label{sec:Lin}
\begin{figure}[b]
\includegraphics[scale=0.52]{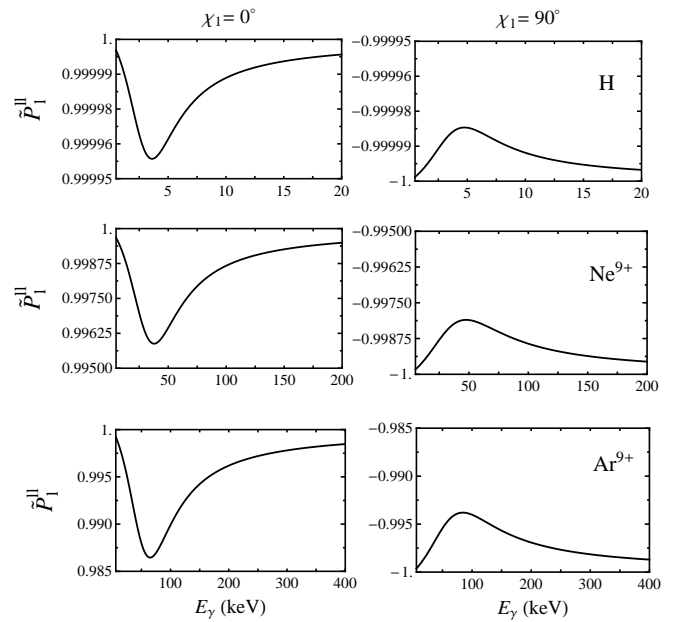}
\caption{First ISP of the scattered light when incident light is linearly polarized ($\tilde{P}^{ll}_1$), as a function of the photon energy. Results are displyed for the  hydrogen (top row), hydrogenlike neon (middle row) and hydrogenlike argon (bottom row) and for the $\hat x$ linearly polarized incident light (left column) and $\hat y$ linearly polarized incident light (right column). }
\label{fig:6}
\end{figure}
Here, we consider the linear polarization scenario as described in Sec. \ref{subsubsec:Lin-scenario}. \Blue{Fig. \ref{fig:6} displays the first ISP, $\tilde{P}_1^{ll}$, for the two incident linear polarizations $\chi_1=0^\circ$ (left column) and  $90^\circ$ (right column). Results are shown for hydrogen (top row), $\textrm{Ne}^{9+}$ (middle row) and $\textrm{Ar}^{17+}$ (bottom row), as a function of the incident photon energy.}  We can see that $\tilde{P}_1^{ll}$ turns out to be $\approx1$ ($-1$) for $\chi_1=0^\circ$ ($90^\circ$) polarization direction, which indicates that polarization-flip (spin-flip) transitions are suppressed (see Eq. \eqref{eq:int_P1_ll}). This indicates that electron-photon spin-spin interaction is weak, since such interaction plays a role in spin-flip transitions. However, the deviations that $|\tilde{P}^{ll}_1|$ has from unity are meaningful for understanding the magnitude of the spin-spin interactions, and at which energy they achieve their maximum. \Blue{In addition, we readily see: i) spin-spin interaction effects are of the order $\sim10^{-3}\, \%$, $\sim\, 0.5 \, \%$ and $\sim\, 2 \, \%$ for $\textrm{H}$, $\textrm{Ne}^{9+}$ and $\textrm{Ar}^{17+}$, respectively; ii) $\tilde{P}_1^{ll}$ has a peak (or a dip) at photon energy value (called $E_\gamma^{max}(Z)$ hereinafter) $\approx 5$ keV, $\approx 50$ keV and $\approx 90$ keV, for $\textrm{H}$, $\textrm{Ne}^{9+}$ and $\textrm{Ar}^{17+}$, respectively.\\
Based on our calculations, we find that the magnetic-dipole term (M1) of the electron-photon interaction operator is mainly responsible for the deviations that $|\tilde{P}^{ll}_1|$ has from unity. Magnetic-dipole M1 can be rigorously shown to be responsible for spin-spin and spin-orbit interactions between the photon and the atomic electron \cite{Filippo:11}: M1$\propto (\vec k\times\epsilon^l_\chi)\cdot (g_s\,\vec S+ \vec L)$, where $\vec S$, $\vec L$ are the spin and orbital angular momentum of the atomic electron, while $g_s$ is the gyromagnetic factor. Since the orbital angular momentum of $1s$ orbital in hydrogenic atoms is zero, M1 here reflects the electron-photon spin-spin interaction. Therefore, we may label the quantity $1-|\tilde{P}^{ll}_1|$ as indicating the magnitude of electron-photon spin-spin interaction. \\
Generally, we find that spin-spin interaction effects are $\propto Z^2$, where $Z$ is the atomic number. In order to have more quantitative idea regarding the energies at which spin-spin interaction is maximized, we recall the multipole expansion in Eq. \eqref{multi-pole decomposition final} where spherical tensor $\bm{a}^p_{LM}(k,\vec r)$ can be expressed in terms of spherical Bessel functions $j_L(kr)$ \cite{L.Safari:12}. The first magnetic contribution, M1, depends on the spherical Bessel function of first order: M1$\propto j_1(kr)$. The first term responsible for the spin-spin interaction in the squared amplitude will thus be $\propto (E1)^* M1\propto j_0(kr) j_1(kr)$. The condition $kr\approx 1.3$ maximizes such term and thus it maximizes the contribution of M1. It is easily seen that such condition, with $r \approx r_A$, where $r_A$ is hydrogenic Bohr radius, is satisfied by energy values $E_\gamma^{max}(Z)$ in Fig. \ref{fig:6}. Moreover, since $r_A\propto Z^{-1}$, the condition $kr_A\approx 1.3$ implies that $E_\gamma^{max}(Z)\propto Z$. This fact is evident from Fig. \ref{fig:6}. We further remark that, since the electron binding energy ($E_b$) in the hydrogenic atoms grows with atomic number as $E_b\propto Z^2$, the ratio $E_\gamma^{max}(Z)/E_b$ decreases as $\propto Z^{-1}$. This implies that, the higher the atomic number is, the closer $E_\gamma^{max}(Z)$ is to the binding energies, which results in enhancing spin-spin interaction effects. In other words, photons which maximize spin-spin interaction in hydrogen atom have energy $E_\gamma^{max}(1)$, which is far away from the atomic binding energy and therefore electron-photon interactions are very weak. As the atomic number grows, $E_\gamma^{max}(Z)$ approaches the atomic binding energy from above and this allows the ion to interact more effectively with photon of such energy resulting in larger spin-spin interaction effects.}

We remark that $\tilde{P}_1^{ll}$ is here equivalent to the degree of linear polarization of the scattered light since, because of symmetry reasons, $\tilde{P}_2^{ll}$ vanishes for all photon energies.

\begin{figure}[t]
\includegraphics[scale=0.6]{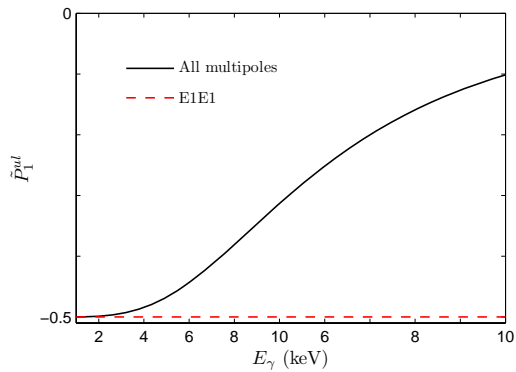}
\caption{(Color online) First ISP of scattered light when incident light is unpolarized ($\tilde{P}_1^{ul}$), as a function of the photon energy. Results are calculated with the account of all photon multipoles (solid-black line) and within the electric dipole approximation (dashed-red line).}
\label{fig:8}
\end{figure}%
Let us consider the case in which incident light is unpolarized and the PDTCS is measured in linear basis. For such a case, the first ISP, $\tilde{P}_1^{ul}$, is displayed as a function of the photon energy in Fig. \ref{fig:8} \Blue{for hydrogen}. As we can see, the first ISP ($\tilde{P}_1^{ul}$) is -0.5 at low photon energies. This means that half of the scattered light is polarized along $y$ direction. On the other hand, such (partial) polarization of the scattered light is almost completely lost at high energies. \Blue{Similar plots for other targets ($\textrm{Ne}^{9+}$ and $\textrm{Ar}^{17+}$), which we do not show here for the sake of simplicity, follow the same behavior as for the hydrogen.} Similar to the previous case, $\tilde{P}_2^{ul}$ vanishes for all photon energies \Blue{and all targets}.

%
%
%
%
\section{Summary and future prospects}
\label{sec:SumConcl}

We studied the polarization characteristics of Rayleigh scattered x-ray photons, based on second-order perturbation theory and Dirac relativistic equation. Experimental scenarios were considered for which light has circular or linear polarization. We derived general analytical expressions for the polarization-dependent total cross sections and integrated Stokes parameters by decomposing the transition amplitude in terms of spherical tensors (angular part) and reduced amplitudes (radial part). Results have been obtained for scattering by \Blue{hydrogen as well as hydrogenlike neon and argon by means of the finite basis-set method based on the Dirac equation. The Stokes parameters of the scattered light were plotted for the photon energies higher than $1s$ ionization threshold of the targets. Special attention has been paid to electron-photon spin-spin interaction.} We found that, circularly polarization of the incident photon, while is totally lost for Rayleigh scattering at low photon energies, is preserved and conveyed to the scattered photon at high photon energies for \Blue{all targets}. We also found that, linear polarization of the incident photon is in general conserved and conveyed to the scattered photon in all energy values. \Blue{However, there is an energy window at which the linear polarization of the scattered photon is slightly jeopardized by the effect of electron-photon spin-spin interactions, being the latter brought by the magnetic-dipole term in the expansion of the electron-photon interaction operator. Such effect can be explored by analyzing the Stokes parameters and it has a contribution from $10^{-3}$ \% to 2 \%, from hydrogen to hydrogenlike argon. Highly energetic photons mostly scatter forwardly. Therefore, the PDTCSs and ISP here theoretically studied may be measured up to a very good approximation by current techniques of measuring angle-differential cross sections.}

%
%
%
%
\section{acknowledgment}
L. S. and F. F. acknowledge financial support by the Research Council for Natural Sciences and Engineering of the Academy of Finland. 
P. A. acknowledges the support of German Research Foundation (DFG) within the Emmy Noether program under Contract No. TA 740 1-1. 
J. P. S. and P. A. acknowledge the support by FCT -- Funda\c{c}\~ao para a Ci\^encia e a Tecnologia (Portugal), through the Projects No. PEstOE/FIS/UI0303/2011 and PTDC/FIS/117606/2010, financed by the European Community Fund FEDER through the COMPETE -- Competitiveness Factors Operational Programme. 
F.F. acknowledges support by Funda\c{c}\~ao de Amparo \`a Pesquisa do estado de Minas Gerais (FAPEMIG) and Conselho Nacional de Desenvolvimento Cient\'ifico e Tecnol\'ogico (CNPq).

%
%
%
%
%


\begin{thebibliography}{36}

\bibitem{G.Weber:2010A} G. Weber, H. Br\"auning, S. Hess \etal, Journal of Instrumentation.{\bf 5}, C07010 (2010).

\bibitem{U.Spillmsnn:2008} U. Spillmann, H. Br\"auning, S. Hess \etal, Rev. Sci. Instum. {\bf 79}, 083101-8 (2008).

\bibitem{D.Protic:2005} D. Protic, T. Stohlker, T. Krings \etal, IEEE Trans. Nucl. Sci. {\bf 52}, 3194-3198 (2005)

\bibitem{C.Spezzani:11} C. Spezzani, E. Allaria, M. Coreno, B. Diviacco, E. Ferrari, G. Geloni, E. Karantzoulis, B. Mahieu, M. Vento, and G. De Ninno, Phys.\ Rev.\ Lett.\ \textbf{107}, 084801 (2011).

\Blue{\bibitem{Hemsing:2013} E. Hemsing, A. Knyazik, M. Dunning \etal, Nature Phys. {\bf 9}, 549 (2013).

\bibitem{Hemsing:2014} E. Hemsing, G. Stupakov, D. Xiang, and A. Zholents, Accepted in Rev. Mod. Phys (2014).}		

\bibitem{F.Fratini:2014}	F. Fratini, L. Safari, A. G. Hayrapetyan, P. Amaro, and J. P. Santos, Europhys. Lett.  {\bf 107}, 13002 (2014). 
						
\bibitem{F.Smend:87} F. Smend, D. Schaupp, H. Czerwinski, M. Schumacher, A. H. Millhouse, and L. Kissel, Phys.\ Rev.\ A {\bf 36} 5189 (1987).


\Blue{\bibitem{Alex:2013}  A. Gumberidze and M. Schwemlein (private communication).

\bibitem{A.Surzhykov:2013} A. Surzhykov, V. A. Yerokhin, T. Jahrsetz, P. Amaro, Th. St\"ohlker, and S. Fritzsche, Phys. Rav. A {\bf 88}, 062515 (2013).}

\bibitem{A.Lewis:2013} A. Lewis, J. Cosmol. Astropart. Phys. {\bf 08}, 053  (2013). doi:10.1088/1475-7516/2013/08/053 

\Blue{\bibitem{PRISM:2014} PRISM, J. Cosmol. Astropart. Phys. {\bf 02} 006 (2014).  doi:10.1088/1475-7516/2014/02/006

\bibitem{J.Trujillo:06} J. Trujillo Bueno, {\it Gentle Introduction to the Physics of Spectral Line Polarization}, in Solar Physics and Solar Eclipses (SPSE 2006), Proceedings of an International Symposium held at Waw an Namos, Libya, 27-29 March 2006, p.77-92 .
http://www.irsol.ch/spse/

\bibitem{E.Landi:04} E. Landi Degl'Innocenti, M. Landolfi, {\it Polarization in Spectral Lines}, (Kluwer Academic Publishers, 2004).

\bibitem{J.O.Stenflo:06} J. O. Stenflo, Astronomical Society of the Pacific Conference Series, Vol. 358, Solar Polarization 4, 215 (2006).

\bibitem{A.J.Dean:08} A. J. Dean, D. J. Clark, J. B. Stephen \etal, Science {\bf 29} 1183 (2008).} 

\bibitem{L.Safari:12_B} L. Safari, P. Amaro, S. Fritzsche, J. P. Santos, S. Tashenov and F. Fratini, Phys.\ Rev.\ A {\bf 86}, 043405 (2012). 

\bibitem{L.Safari:12} L. Safari, P. Amaro, S. Fritzsche, J. P. Santos, and F. Fratini, Phys.\ Rev.\ A {\bf 85}, 043406 (2012).

\bibitem{Filippo:11} F.~Fratini, \textit{One- and Two-photon decays in Atoms and Ions} (LAP, Saarbr\"ucken, Germany, 2011).

\bibitem{Akhiezer:65} A.~I.~Akhiezer and V.~B~Berestetskii, {\it Quantum Electrodynamics} (Wiley, New York, 1965).

\bibitem{M.E.Rose:57} M.~E.~Rose, \textit{Elementary Theory of Angular Momentum} (John Wiley, New York, 1957).

\bibitem{Sakurai:94} J.~J.~Sakurai, \textit{Modern Quantum Mechanics} (Addison-Wesley, 1994).

\bibitem{S.P.Goldman:81} S.~P.~Goldman and G.~W.~F.~Drake, Phys.\ Rev.\ A \ \textbf{24}, 183 (1981).

\bibitem{F.Fratini-Hyrapetyan:11} F. Fratini, and A. G. Hayrapetyan, Phys.\ Scripta.\ {\bf 84}, 035008 (2011).

\bibitem{Edmond:eq.4.5.1} A. R. Edmond, {\it Angular momentum in quantum mechanics} (Princeton University Press, 1957), equation 4.5.1 .

\bibitem{saj1996} J.~Sapirstein and W.~R.~Johnson, J.\ Phys.\ B\ \textbf{29}, 5213 (1996).

\bibitem{jbs1998} W.~R.~Johnson, S.~A.~Blundell and J.~Sapirstein, Phys. Rev. A \textbf{37}, 307 (1988).

\bibitem{ZatBar2008} O.~Zatsarinny and K.~Bartschat, Phys. Rev. A {\bf 77}, 062701 (2008).

\bibitem{FroZat2009} C.~Froese~Fischer and O.~Zatsarinny, CPC {\bf 180}, 879 (2009).

\bibitem{sfi1998} J.~P.~Santos, F.~Parente and P.~Indelicato, Eur.\ Phys.\ J.\ D\ \textbf{3}, 43 (1998).

\bibitem{P.M.D.G.Amaro} P. M. D. G. Amaro, \textit{Study of Forbidden transitions in Atomic Systems} (FCT-NUL and UPMC, 2013).


\Blue{\bibitem{F.Fratini:11_2} F.~Fratini, M.~C.~Tichy, Th.~Jahrsetz, A.~Buchleitner, S.~Fritzsche, and A.~Surzhykov, Phys. Rev. A {\bf 83}, 032506 (2011).}

\bibitem{F.Fratini-S.Trotsenko:11} F. Fratini, S. Trotsenko, S. Tashenov, Th. St\"ohlker and A. Surzhykov, Phys. Rev. A {\bf 83}, 052505 (2011).

\Blue{\bibitem{P.Amaro:2012} P. Amaro, F. Fratini, S. Fritzsche, P. Indelicato, J. P. Santos, and A. Surzhykov, Rev. A {\bf 86},  042509 (2012).}

\bibitem{A.Surzhykov:10} A. Surzhykov, A. Volotka, F. Fratini, J. P. Santos, P. Indelicato, G. Plunien, Th. St\"ohlker and S. Fritzsche, Phys. Rev. A \textbf{81}, 042510 (2010). 

\bibitem{PolPol} F.~Fratini and A.~Surzhykov, Hyp. Int. {\bf 199}, 85 (2011).

\bibitem{L.Safari:2014} L. Safari, P. Amaro, J. P. Santos, and F. Fratini, Phys. Rev. A {\bf 90}, 014502 (2014).


\end{thebibliography}
\end{document}